# Geometry of a Centrosymmetric Electric Charge

Yu. A. Khlestkov

*Moscow Engineering Physics Institute, Moscow, 115409 Russia*

*e-mail: khlestkov@yandex.ru*

**Abstract**

The gravitational description given for an electric charge qn the basis of exact solution of the Einstein–Maxwell equations eliminates Coulomb divergence. The internal pulsating semiconfined world formed by neutral dust is smoothly joined with parallel Reissner–Nordstrem vacuum worlds via two static bottlenecks. The charge, rest mass, and electric field are expressed in terms of the space curvatures. The internal and external parameters of the maximon, electron, and the universe form a power series.

## 1. INTRODUCTION

Elementary particles are treated as point objects in planar spacetime. Their internal structure can be considered in the space of the general theory of relativity (GTR), whose curvature is equivalent to the gravitational field [1].

Initially, Einstein planned, first, to geometrize physical fields and their sources and, second, to represent the discreteness of space-time as well as parameters of elementary particles, their nonlocalizability and stochastic behavior (quantum effects) as manifestations of properties of a "continuous" gravitational field.

This study provides a simple example illustrating the possibility of implementation of the first part of this program. The internal structure of an electric charge is described on the basis of the exact solution to the Einstein–Maxwell equations for a nonstationary centrosymmetric dust-like matter (dust) and an electric field [2].

It should be noted that the second part of the program has not been realized as yet; i.e., the quantum behavior of elementary particles has neither been described in a nonlinear geometrized theory nor interpreted, and the spin has not been taken into account; the solution obtained for an electric charge cannot be extended to real charged elementary particles (electron, proton, etc.). We will only consider the description of geometry of an abstract electric charge, whose charge and rest mass coincide with the corresponding quantum numbers of the real electron, maximon, universe, etc.

This solution clarifies the meaning of two universal constants, viz., electric charge $e$ and rest mass $m_0$, which are found to be first integrals of the Einstein–Maxwell equations (integrals of motion) and which can be determined from the space-time curvatures at any point of the space-time.

The procedure in which a source of gravitational field (energy-momentum tensor of matter appearing on the right-hand side of the equations) must be specified "manually" has been usually regarded as poor since Einstein's time. For this reason, the possibility of complete geometrization of physical fields is considered as dubious.

The solution obtained here shows that this is not quite correct. We specify only the "filling" of the system to be studied (matter, physical fields and interactions, the presence of charges, pressure, rotation, torsion, etc.). Then, in view of the nonlinearity and selfconsistency of the Einstein system of equations and equations for potentials of physical fields, all physical parameters of the system (densities, velocities, and field strengths) are expressed in terms of geometry as a result of the solution of such a system of equations.

In contrast to the linear situation, which is responsible in al probability for such an attitude to the Einstein equations with a right-hand side, it is impossible in the general case to characterize the properties of the functions satisfying the nonlinear equations prior to their solution in view of the absence of a superposition principle. A nonlinear equation has no "source" or the "right- and left-hand sides" in the linear sense.

It is also equally important that the space-time curvature eliminates the main disadvantage of theories in the Minkowski space-time, viz., Coulomb divergence of the field of a point charge, which generates divergences in the existing quantum theories.

Strictly speaking, a planar space must be empty, since if the Riemann-Christoffel curvature tensor is zero, the conservative

Einstein tensor must also be zero, as well as the energy-momentum tensor of matter and, hence, the densities and potentials of physical fields.

In addition, this solution at last clarifies the reason for the prevailing concept of negligibility of gravitational effects over the classical length (due to extremely small value of the gravitational radius as compared to the classical radius), as well as the idea that gravitational interaction (space-time curvature) become significant either over limiting lengths in microworld (such as Planck's length), or over the scale of the universe in macro- and megaworlds.

The reason appears as paradoxical: such a situation in vacuum (vanishing smallness of the "potential gravitational field" as compared to the potential electric field of an isolated stationary point charge, $\sqrt{k}m_0/r^2 \ll e/r^2$, where $k$ is the gravitational constant) is explained just by the large "gravitational mass defect" due to the strong gravitational field (large space-time curvature) inside the charge, which reduces the energy of charge in vacuum observed from outside (from the "Newtonian" point of view, the gravitation field possesses a sort of "negative energy").

The gravitational interaction, in accordance with its definition, is universal and is manifested over any length, mapping physical fields on geometry [1].

A brief review of the literature concerning this problem in GTO and the reason for which a rigorous solution of the problem of the source of electric charge and rest mass of a particle could not be obtained earlier are given in the Appendix.

## 2. FORMULATION OF THE PROBLEM

Let us suppose that the gravitational field in a centrosymmetric orthogonal nonstationary metric [1] (in the $\tau$, $r$, $\theta$, $\varphi$ coordinates)

$$ds^2 = e^\nu d\tau^2 - e^\lambda dr^2 - R^2(\tau, r)d\sigma^2, \quad (1)$$
$$d\sigma^2 = d\theta^2 + \sin^2\theta \, d\varphi^2$$

in a synchronous co-moving frame of reference is defined by the energy-momentum tensor, whose mixed components are given by

$$diag(\varepsilon_s + \varepsilon_f, \varepsilon_f, -\varepsilon_f, -\varepsilon_f),$$

which corresponds to an ideal dust-like generally charged matter with a charge density $\rho_f$, energy density $\varepsilon_s(\tau, r)$, and electromagnetic field with energy density $\varepsilon_f(\tau, r)$.

The system of the Einstein-Maxwell equations in the given case can be written in the form

$$\Phi = \frac{4\pi}{\kappa} R\left(1 - e^{-\lambda}R'^2 + e^{-\nu}\dot{R}^2\right),$$
$$\dot{\Phi} = 4\pi R^2 \dot{R}\varepsilon_f,$$
$$\Phi' = 4\pi R^2 R'\left(\varepsilon_f + \varepsilon_s\right),$$
$$\dot{\varepsilon}_f + 4\dot{R}\varepsilon_f/R = 0,$$
$$\varepsilon'_f + 4R'\varepsilon_f/R = \sqrt{8\pi\varepsilon_f}\rho_f e^{\lambda/2},$$
$$\dot{\varepsilon}_s + \left(\dot{\lambda} + 4\dot{R}/R\right)\varepsilon_s/2 = 0,$$
$$\dot{\rho}_f + \left(\dot{\lambda} + 4\dot{R}/R\right)\rho_f/2 = 0$$

(we assume that the cosmological term is zero). Here, $\dot{R} = \partial R/\partial \tau$, and $\kappa = 8\pi k/c^4$ is the Einstein constant.

The general solution to the Cauchy problem depends on three arbitrary functions of $r$, viz., integrals of motion corresponding to the initial distribution of energy density $\varepsilon_s(0, r)$ of matter, charge density $\rho_f(0, r)$, and radial velocity $\beta_r(0, r)$, which is defined for $\tau = 0$.

For such functions, we can choose [2] the electric charge in the form

$$Q(r) = 4\pi \int \rho_f e^{\lambda/2} R^2 dr + Q_0,$$

the total energy of matter in the form

$$E(r) = 4\pi \int (\varepsilon_s + \varepsilon_f) R^2 R' dr + Q^2/2R + E_0,$$

and the function

$$f(r) = e^{-\lambda/2} R' + qQ/R,$$

where $Q_0$ and $E_0$ are constants and $q = \rho_f/\varepsilon_s$.

In terms of these quantities, we can express the energy densities of matter and field [2]:

$$\varepsilon_s = \frac{E'}{4\pi R^2 R'\left(1 - qQe^{\lambda/2}/RR'\right)}, \varepsilon_f = \frac{Q^2}{8\pi R^4}.$$

The solution to the Einstein-Maxwell equations is obtained in the cases when one of the arbitrary functions of $r$ becomes constant [2]. There exist three types of solutions,

$$Q = Q_0, \quad R_g = R_{g0}, \quad R_f = R_{f0},$$

where

$$R_g = \kappa E/4\pi$$

is the gravitational radius,

$$R_f = Q^2/2E = R_c^2/R_g$$

is the classical (electromagnetic) radius, and

$$R_c = \sqrt{k}Q/c^2$$

is the so-called critical radius.

In the given problem, we are interested in the first type of solutions with a constant electric charge $Q_0$. In this case, $\rho_f = 0$; i.e., the dust is neutral. An interesting situation arises: charge $Q_0$ is an integral of motion of a neutral gravitating matter and, in turn, generates an electromagnetic field with energy density $\varepsilon_f$ and field strength $E_r = \pm\sqrt{8\pi\varepsilon_f}$ of the radial electric field in the co-moving reference frame (or is generated by this field and the neutral matter).

It would be interesting to find out what are the inner and outer (vacuum) worlds of the object called the electric charge and to express its physical parameters (charge, mass, and radius) in terms of the curvature of the pseudo-Riemannian space of the inner world. Let us prove that even in the simplest centrosymmetric case in the absence of the inevitable rotation, the charge is a non-Euclidean topological construction, viz., a semiconfined pulsating world with two static extremal surfaces (bottlenecks) in the spacetime.

## 3. INNER WORLD

We choose an attractive (from the standpoint of physics) solution for $Q = Q_0$, which corresponds to a semiconfined time-periodic world [2]

$$e^\nu = 1, \quad e^\lambda = R'^2/f^2,$$
$$f^2 < 1, \quad 4R_f(1-f^2)/R_g < 1,$$
$$R = \frac{R_g}{2(1-f^2)}(1-\delta\cos\eta), \quad (2)$$
$$\tau - \tau_r = \frac{R_g}{2(1-f^2)^{3/2}}(\eta - \delta\sin\eta),$$

where

$$\delta = \sqrt{1 - 4R_f(1-f^2)/R_g}$$

and $\tau_r(r)$ is an arbitrary function of $r$, which is determined by the method for measuring time in the congruence of observers.

For $Q_0 = 0$, this solution is transformed to the well-known nonstationary metrics of the Tolman–Friedmann confined world of neutral dust [1]; in the absence of matter the solution is transformed into the Reissner–Nordstrem static world of a solitary charge, which in turn is transformed to the Schwarzschild world of a point mass.

A remarkable property of metric (2) is the absence of point-like singularity of the type of an infinite Gaussian curvature of radial spheres: since $\delta < 1$, the radius $R(\tau, r)$ of the internal scalar curvature of the 2-surface ($\tau$ = const, $r$ = const) never vanishes anywhere ($R(\tau, r) \neq 0$) if the integral of motion $R_g(r)$ is defined appropriately.

Since $E_r = Q_0/R^2$, gravitation (i.e., the spacetime curvature) removes the Coulomb divergence of the classical point charge field in the Minkowski spacetime.

Let us specify the initial conditions for $\tau = 0$. Let us suppose that the density of matter in the state of maximal expansion of the inner world is constant:

$$\eta = \pi, \quad \dot{R}(0) = 0, \quad \varepsilon_s(0) = \varepsilon_0.$$

Here, zero in the parentheses indicates the dependence of quantities on $r$ for $\tau = 0$. Integrating the relation

$$R'_g = \kappa R^2 R' \varepsilon_s$$

for $\tau = 0$, we obtain the relation between the total energy and the density of matter in the initial state,

$$R_g = \frac{\kappa\varepsilon_0}{3}R^3(0) + C_g \quad (3)$$

where $C_g$ is the integration constant. All quantities can now be expressed in terms of $R(0)$ and $R$ (we set $C_g = 0$):

$$R_g = \frac{\kappa\varepsilon_0}{3}R^3(0), \quad R_f = \frac{3R_c^2}{\kappa\varepsilon_0}\frac{1}{R^3(0)},$$
$$f^2 = 1 - \frac{R_g}{R(0)} + \frac{R_c^2}{R^2(0)},$$
$$\varepsilon_s = \varepsilon_0\frac{R^2(0)R'(0)}{R^2 R'}, \quad \varepsilon_f = \frac{R_c^2}{\kappa R^4}.$$

## 4. STATIC SURFACE

Let us define the object, viz., a static 2-surface ($h$)

$$r = r_h, \quad R = R_h, \quad \dot{R}_h = 0.$$

The static conditions ($\dot{R}_h = 0$) lead to the conditions

$$\delta_h = 0, \quad 1 - f_h^2 = \frac{R_{gh}^2}{4R_c^2};$$

substituting these conditions into Eq. (2), we find that the scalar curvature radius of a static 2-surface is always equal to doubled classical radius,

$$R_h = 2R_{fh},$$

i.e., to the ratio $R_h = Q_0^2 / E_h$ of the squared charge to the total energy of the inner world for $r = r_h$. For $Q_0 \neq 0$, this radius always differs from zero.

All parameters on the given surface can be expressed in terms of critical radius $R_c$ and dimensionless parameter $\xi$:

$$R_h = \xi R_c = \frac{R_{gh}}{2}\xi^2, \quad R_{gh} = \frac{2R_c}{\xi},$$
$$1 - f_h^2 = \frac{1}{\xi^2}. \tag{4}$$

Since $f_h^2 < 1$, parameter $\xi > 1$; consequently, the radius of curvature of the static sphere is larger than the critical radius ($R_h > R_c$) for any charge.

The extremal object for which $\xi = 1$ is called a maximon. It is the only static 3-object with a constant curvature and a constant density, with a spherical system of coordinates (which is nondegenerate on its surface), with the minimal radius, and with the maximal gravitational radius.

In solving the Cauchy problem, we choose the initial conditions so that $R(0) \geq R_h$ (i.e., the radius of curvature of the static sphere has its minimal value in the initial state of maximal expansion. In this case, for the integrals of motion we have

$$R_g = R_{gh} \frac{R^3(0)}{R_h^3} \geq R_{gh},$$
$$R_f = R_{fh} \frac{R_h^3}{R^3(0)} \leq R_{fh};$$

i.e., the total energy of the inner world on the static surface is minimal, while the electromagnetic field energy is maximal.

Let us define the "rest mass" $m_0$ as the total mass (energy) of the inner world on the static sphere,

$$R_{gh} = \frac{2k\mathrm{E}_h}{c^4} = \frac{2km_0}{c^2}, \quad m_0 = \frac{m_c}{\xi},$$

where $m_c = Q_0 / \sqrt{k}$ is the maximon mass.

Applying relation (3) to the static sphere and taking into account relation (4) between its radius and the gravitational radius, we find that the dust energy density in the initial state (and, hence, on the given sphere) is finite and unambiguously determined by its parameters,

$$\varepsilon_0 = \varepsilon_c / \xi^4$$

where $\varepsilon_c = 6 / \kappa R_c^2$ is the critical energy density.

## 5. MAXIMAL EXPANSION STATE

The inner world of a charge in the initial state $\eta = \pi$ can be described by the differential equation

$$e^{-\lambda(0)}(R^2(0)')^2 = 4\left[R^2(0) - \frac{2}{\xi^2 R_h^2}R^4(0) + R_c^2\right], \tag{5}$$

whose solution can be written in the form

$$R^2(0) = 2a_0^2(1 - \delta_0 \cos 2\chi),$$
$$\int e^{\lambda(0)/2} dr = 2a_0 \chi, \tag{6}$$

where the dimensionless coordinate $\chi \in [0, \pi]$ [1],

$$\delta_0 = \sqrt{1 + 8/\xi^4}, \quad 2a_0 = \frac{\xi R_h}{\sqrt{2}} = \sqrt{\frac{3}{\kappa \varepsilon_0}},$$

and $e^{\lambda(0)}$ is a function of $r$, which depends on the method for measuring radial lengths.

Solution (6) implies that a semi-confined world has two static spheres (geometrical images of charges with opposite polarities, i.e., charged particles-antiparticles), on which the radial electric field has opposite directions. These spheres are located at points $\chi_h$ and $\pi - \chi_h$, where

$$\chi_h = \frac{1}{2} \arccos \frac{1 - 4/\xi^2}{\delta_0}$$

The given solution for $R_c = 0$ is transformed into the well-known solution for the Tolman world [1]. The maximal *radial length* of the inner world in the confined model is given by

$$l_r(\pi) = \int e^{\lambda(0)/2} dr = 2\pi a_0$$

Thus, the radial length of the inner world in the maximal expansion state is defined, in accordance with relation (6), by the energy density of the matter in the initial state.

For $\eta = \pi$ and $\chi = \pi/2$, quantities $R(0)$ and $R_g$ assume their maximal values, such that $R_{max} \approx R_{g\,max}$:

$$R_{max} = \frac{\xi R_h}{2}\sqrt{1+\delta_0} = \sqrt{2}a_0\sqrt{1+\delta_0},$$

$$R_{g\,max} = \frac{\xi R_h}{4}(1+\delta_0)^{3/2} = \frac{a_0}{\sqrt{2}}(1+\delta_0)^{3/2}.$$

The maximal value of the gravitational radius (total energy) can be juxtaposed to the total mass $M$ of the inner world in the maximal expansion state:

$$R_{g\,max} = 2kM/c^2.$$

## 6. GEOMETRIZATION OF THE CHARGE

The curvature ${}_\alpha K^{(a)}_\beta$ of the 2-surface $S^{(2)}$ formed by the coordinate lines $\{x^\mu, x^\nu\}$ is perpendicular to coordinates $x^\alpha, x^\beta$, $\alpha \neq \beta \neq \mu \neq \nu$, and observed from the measurement space $a$ ($a = 2, 3, 4, \ldots$) can be expressed in terms of the Riemann-Christoffel tensor of the corresponding space and the modulus of the metric on the surface [1]:

$$_\alpha K^{(a)}_\beta = R^{(a)}_{\mu\nu\mu\nu}/(g_{\mu\mu}g_{\nu\nu} - g_{\mu\nu}^2) \quad (7)$$

(summation over the indices is not envisaged). In metric (1), we obtain from expression (7)

$$_0 K^{(2)}_r = \frac{1}{R^2}, \quad _0 K^{(3)}_r = \frac{1}{R^2}(1 - e^{-\lambda}R'^2),$$
$$_0 K^{(4)}_r = \frac{1}{R^2}(1 - e^{-\lambda}R'^2 + e^{-\nu}\dot{R}^2), \quad (8)$$

$$_0 K^{(2)}_\varphi = _0 K^{(3)}_\varphi, \quad _0 K^{(3)}_\varphi = \frac{1}{R^2 r'}(1 - e^{-\lambda}R'^2)',$$
$$_0 K^{(4)}_\varphi = \frac{1}{R^2 r'}(1 - e^{-\lambda}R'^2 + e^{-\nu}\dot{R}^2)' \quad (9)$$

$$_0 K^{(a)}_\theta = _0 K^{(a)}_\varphi.$$

The curvature of the 3-hypersurface $S^{(3)}$ orthogonal to coordinate $x^\alpha$ is equal to the sum of curvatures (7) over index $\beta$:

$$_\alpha K^{(a)} = _\alpha K^{(a)}_\beta + _\alpha K^{(a)}_\gamma + _\alpha K^{(a)}_\delta$$
$$\alpha \neq \beta \neq \gamma \neq \delta. \quad (10)$$

The sum of 4-curvatures of all area elements orthogonal to the $x^0$ axis is equal to the $G^0_0$ component of the Einstein tensor,

$$_0 K^{(4)} = _0 K^{(4)}_r + 2\,_0 K^{(4)}_\varphi = G^0_0 = \kappa(\varepsilon_f + \varepsilon_s). \quad (11)$$

The sum of 4-curvatures of all area elements orthogonal to the $x^1$ axis is equal to the $G^1_1$ component of the Einstein tensor,

$$_r K^{(4)} = _r K^{(4)}_0 + 2\,_r K^{(4)}_\varphi = G^1_1 = \kappa(\varepsilon_f - p_s),$$

where $p_s$ is the pressure of the matter (which differs from zero in the general case).

The scalar curvature of the 4-space (Gaussian, or internal, curvature [1]) is equal to the sum of all curvatures (10) orthogonal to the axes $x^0$, $x^1$, $x^2$, and $x^3$; in the present case, it is given by

$$K^{(4)} = _0 K^{(4)} + _r K^{(4)} + 2\,_\varphi K^{(4)}$$
$$= G = G^0_0 + G^1_1 + 2G^2_2 = \kappa\varepsilon_s = R'_g/R^2 R' \quad (12)$$

The Einstein-Maxwell equations lead to the following relation between curvatures (7)-(12) and physical characteristics:

$$\kappa\varepsilon_s = K^{(4)},$$
$$\kappa\varepsilon_f = _0 K^{(4)} - K^{(4)},$$
$$1 - f^2 = _0 K^{(2)-1}_r\,_0 K^{(3)}_r, \quad (13)$$
$$R_c = _0 K^{(2)-1}_r (_0 K^{(4)} - K^{(4)})^{1/2},$$
$$R_g = _0 K^{(2)-3/2}_r (_0 K^{(4)} - K^{(4)} + _0 K^{(4)}_r).$$

It was noted above that $_0 K^{(a)}_r \neq \infty$, i.e., singularity of radial spheres is absent ($\varepsilon_f \neq \infty$).

Relations (13) make it possible to express two fundamental constants (charge $e$ and rest mass $m_0$), which are the first integrals of the given gravitating system ($m_0$ is equal to the total mass $M(r)$ of the inner world on the static sphere $r = r_h$) in terms of curvatures and two other constants, viz., $c$ (corresponding to the presence of a light cone for the chosen signature) and $k$ (relating the geometry to the physics):

$$e = \frac{c^2}{\sqrt{k}}\left[_0 K^{(2)-1}_r (_0 K^{(4)} - K^{(4)})^{1/2}\right],$$
$$m_0 = \frac{c^2}{2k}\left[_0 K^{(2)-3/2}_r (_0 K^{(4)} - K^{(4)} + _0 K^{(4)}_r)\right] \quad (14)$$

To find constant $e$, we can determine the curvatures at any point of the inner and outer vacuum world of the electric charge ($K^{(4)} = 0$ in vacuum). Constant $m_0$ can be sought from the curvatures at any point of the vacuum world and on the static sphere. Inside the charge, the mass will be a function of $r$.

The electromagnetic field of a charge that is at rest in vacuum, which is represented outside and inside by the radial electric field in a reference frame co-moving with the dust-like matter, can also be expressed in the entire space in terms of its curvatures:

$$E_r = \frac{c^2}{\sqrt{k}}(_0K^{(4)} - K^{(4)})^{1/2}.$$

Thus, an interesting possibility of experimental determination of physical parameters of objects from measuring geometrical quantities; for fundamental constants, the same values will be obtained at any point in space.

### 7. BOTTLENECK

Let us define a bottleneck in the space-time as a 2-surface of extremal curvature. In the simplest centrosymmetric case, we can speak of the bottleneck as an extremal surface of radial spheres, which is orthogonal to time and radial coordinates. If the bottleneck is static in the co-moving reference frame (i.e., its curvature does not change with time and it does not move along the radial coordinate), it coincides with static sphere and the conditions for its existence have the form

$$_0K_{rh}^{(4)} > 0, \qquad _0K_{rh}^{(4)'} = 0, \qquad _0K_{rh}^{(4)\bullet} = 0 \quad (15)$$

The bottleneck will display the maximal curvature of radial spheres ($\eta = \pi$) or inflection for $_0K_{rh}^{(4)''} \leq 0$, and the minimal curvature for $_0K_{rh}^{(0)''} > 0$ ($\eta = 0$). Substituting expressions (8) for curvatures into conditions (15), we obtain

$$_0K_r^{(4)'} = R^{2'} \frac{_0K_\varphi^{(4)} - _0K_r^{(4)}}{R^2}$$
$$= \frac{R'}{R^4}\left(\frac{4R_c^2}{R} - 3R_g + \kappa\varepsilon_s R^3\right) = 0. \quad (16)$$

$$_0K_r^{(4)\bullet} = \frac{R^\bullet}{R^4}\left(\frac{4R_c^2}{R} - 3R_g\right) = 0,$$
(17)

which means that a static bottleneck exists for $R_h^\bullet = 0$ and either for

$$\kappa\varepsilon_0 R_h^3 = 3R_{gh} - 4R_c^2 / R_h$$

or for $R_h^{'} = 0$. In the former case, the quantity $C_g$ in expression (3) for the gravitational radius cannot be set equal to zero any longer; $R(0)$ as a solution to Eq. (5) will be expressed in terms of elliptical functions. Let us consider the second condition $R_h^{'} = 0$.

Since $e^{\lambda_h} = R'^2 / f^2 = 0$ for $\xi \neq 1$, the determinant of the metric tensor vanishes at the bottleneck. Consequently, the spherical system of coordinates degenerates on it. However, all its geometrical parameters (curvatures) and the corresponding physical quantities (mass, dust density, electromagnetic field energy density, and field strength) are finite; i.e., this singularity is of purely coordinate nature. It is not reflected in physics or geometry in any way.

It should also be noted that function $e^{\lambda(0)}$ in solution (6) must vanish on the bottleneck in this case.

### 8. JOINING WITH VACUUM

Solution (2) covers the entire space-time and does not require any supplements. Nevertheless, using static bottlenecks, the inner semi-confined world (2) can be smoothly continued to two Reissner–Nordstrem vacuum worlds. In the curvature coordinates, we have

$$R_g^{'} = 0, \qquad R^\bullet = 0, \qquad R' = 1, \qquad R = r,$$
$$\varepsilon_s = 0, \qquad e^\nu = e^{-\lambda} = A(r),$$
$$ds^2 = A(r)d\tau^2 - \frac{dr^2}{A(r)} - r^2 d\sigma^2, \qquad (18)$$
$$A(r) = 1 - \frac{r_g}{r} + \frac{r_c^2}{r^2}$$

where

$$r_g = R_{gh} = \frac{2km_0}{c^2}, r_c = R_c = \frac{Q_0\sqrt{k}}{c^2}.$$

Since joining is carried out over the bottleneck whose curvature is extremal from inside and on which the metric has a coordinate singularity $g_{11} = 0$, metric (18) must be transformed to another radial coordinate $\tilde{r}$ [1] to nullify the metric coefficient $g_{\tilde{1}\tilde{1}}$ on the 2-surface of joining $r = 2r_f$ as in the case of the internal solution. Naturally, the transformation Jacobian $J = r_{,\tilde{r}}$ also vanishes on this surface (e.g., $r = \tilde{r} + 4r_f^2/(2r_f + \tilde{r})$, where $r_f = R_{fh} = Q_0^2/2E_h^2$). Direct substitution of an arbitrary transformation $r = r(\tilde{r})$ into the Einstein equations [1] under the condition $J_h = 0$ readily shows that transformed metric (18), which is independent of the world time, satisfies these equations on the bottleneck.

At the bottleneck, $R'_h = r_{,\tilde{r}} = 0$ and the quantity $G_1^1$ (mixed component of the Einstein conservative tensor also turns out to be continuous, which is physically equivalent to the continuity of the electric field upon a transition to vacuum. Component $G_0^0$ of this tensor experiences a first-kind discontinuity, which corresponds to a sharp dust-vacuum interface from the standpoint of physics.

The joining procedure satisfies the Likhnerovich conditions: if $f(x^\mu) = 0$ (equation for the joining surface, i.e., the equation $r - r_h = 0$ in the present case), product $G_\mu^\nu f_{,\nu}$ is found to be continuous. In fact, cutting a part $r < 2r_f$ from the Reissner–Nordstrem metric, we discard the singularity $r = 0$ inherent in the vacuum solution and obtain an extended (bulk) material Reissner–Nordstrem field source.

Figure 1 shows qualitatively the hierarchy of spaces for various simple field sources, where the threedimensional hypersurface (observed physical space) is represented for better visualization by a curve along which coordinate $r$ changes (i.e., the cross section of a 2D surface of revolution with one of cyclic coordinates, θ or $\varphi$, varying along its second direction). The distance from the rotational axis is proportional to radius $R(r)$ of the 2D Gaussian curvature; the convexity or concavity of the surface depends on the signs of curvatures $_0K_r^{(3)}$ and $_0K_\varphi^{(3)}$. It can be seen from Fig. 1 that the given field had to be slightly deformed by a transformation (dotted curve) for passing from a point-like to a bulk source in the Reissner–Nordstrem metric and for its smooth joining with the internal solution.

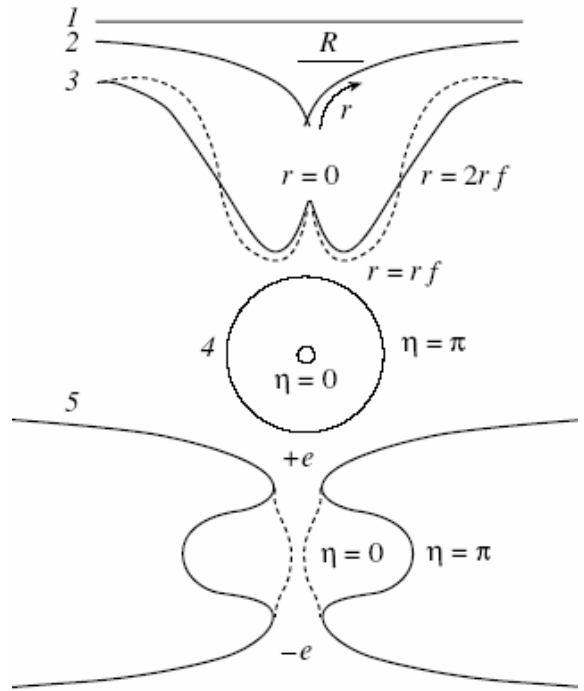

Fig. 1. Geometry of centrosymmetric metrics: planar empty Minkowski space-time (1); Schwarzschild solution for the vacuum field of point mass $m_0$ (2); Reissner–Nordstrem solution for the vacuum field of point charge $e$ having a rest mass of $m_0$ (dotted curves show the geometry of the transformed metric with a Gaussian curvature extremum for $r = 2r_f$) (3); Tolman–Friedmann solution for a confined world of dust-like matter in the maximal expansion state (η = π), having a singularity in the maximal compression state (η = 0) (4); solution for the inner world of an electric charge consisting of neutral dust and a radial electric field pulsating from the maximal expansion state (η = π) joined with two parallel vacuum Reissner–Nordstrem worlds via two static bottlenecks (charges +e, –e) (5).

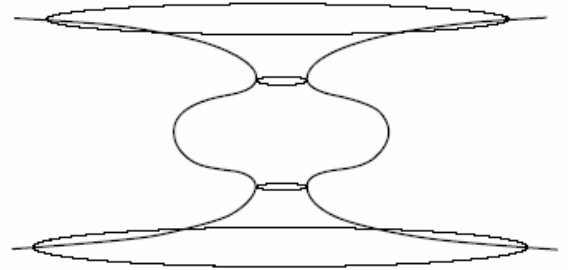

Fig. 2. Geometry of the inner and outer worlds of an electric charge: the physical 3-hypersurface is simplified to a 2-surface of revolution, which is formed by the radial and one of cyclic coordinates. This surface is connected with parallel vacuum world and an antiworld via two bottlenecks (equivalent to a particle and an antiparticle with opposite charges).

Thus, the electric charge appears geometrically as two parallel vacuum spaces in which tensions are equivalent to an electric field and which are connected via two static bottlenecks with a radius equal to doubled classical radius. Between these bottlenecks, a nonstationary confined world is situated, in which tensions are equivalent to a neutral dust-like matter and an electric field and which pulsates from the maximal expansion state to the maximal compression state. This pattern is shown qualitatively in Fig. 2 in the form of a 2-surface of revolution in accordance with the same rules as the curves in Fig. 1. Quantity $R(\eta, \chi)$ does not vanish anywhere. The strength of the radial electric field attains its maximal value on the bottlenecks and decreases in inverse proportion to the square of the Gaussian curvature radius with increasing distance from the bottlenecks to the inner or outer space of the charge.

The only divergence preserved in view of central symmetry and nonstationarity of the metric is the kink on the 2-surface formed by coordinates $\{r, \theta\}$ or $\{r, \varphi\}$ when expansion is replaced by compression. The kink moves according to a definite law $\chi(\eta)$ (singularity of the type $R' = 0$, $R_g{}' \neq 0$, $_0 K_\varphi^{(a)} \to \infty$) and corresponds to an infinitely large value of the energy of the matter ($\varepsilon_s \to \infty$). However, this singularity is immaterial; it appears due to simplicity of the model and is absent on the bottleneck.

## 9. EXTERNAL AND INTERNAL PARAMETERS OF THE CHARGE

Let us suppose that the electric charge of the world is equal to the fundamental charge ($Q_0 = e$) and that constants $c$, $k$, and $e$ have experimental values. Then the parameters of the world on the bottleneck and inside the charge are determined by the value of $\xi$. The maximon ($\xi = 1$) is an object with the critical parameters

$$f_h = 0, \; \delta = 0, \; \delta_0 = 3, R^\bullet = R' = 0,$$
$$R_h = 2R_{fh} = R_{gh}/2 = R_c = R_{max}$$
$$= R_{min} = 2\sqrt{2} a_0, \quad (19)$$
$$m_0 = M = m_c = \frac{e}{\sqrt{k}}, \; \varepsilon_0 = \frac{6}{\kappa R_c^2} = \varepsilon_c,$$
$$R_c = \frac{km_c}{c^2}.$$

Here, $R_{min}$ is the radius of curvature for $\eta = 0$ and $\chi = \pi/2$, i.e., in the state of maximal compression. The parameters of the remaining objects can be expressed in terms of the critical parameters (19):

external,

$$R_h = \xi R_c, \qquad m_0 = m_c / \xi, \quad (20a)$$

and internal,

$$R_{max} = \frac{\xi^2}{2} \sqrt{1+\delta_0} R_c,$$
$$M = \frac{\xi^2}{8}(1+\delta_0)^{3/2} m_c, \quad (20b)$$
$$\varepsilon_0 = \rho_0 c^2 = \varepsilon_c / \xi^4.$$

In this model, the value of $\xi$ is not bounded from above. An important factor is that the experimental value of $\xi$ for the electron is known exactly from its external parameters (20a):

$$\xi_e = e/\sqrt{k} m_0 = 2{,}04 \cdot 10^{21}$$

(it should be recalled that we are speaking here of an abstract object with a charge and a rest mass of a real elementary particle, electron, naturally disregarding rotation in the simplest case of central symmetry). The experimental estimate of the value of $\xi$ for our universe is also known, but now from its internal parameters (20b), viz., the total mass and radius, under the assumption that a closed cosmological model is applicable [1]:

$$\xi_u \approx (2\sqrt{2} M / m_c)^{1/2} \approx (\sqrt{2} R_{max} / R_c)^{1/2}$$

($\delta_0 \approx 1$ for the Universe).

We assume that the electric charge of the Universe is also equal to the fundamental charge.

Consequently, the internal parameters of the electron and, accordingly, the external parameter of our universe are known exactly. The result is curious: the electron contains a universe whose mass in the state of maximal expansion is $M = 2.73 \times 10^{36}$ g (i.e., on the order of $10^3$ masses of the Sun) and whose radius $R_{max} = 4.06 \times 10^8$ cm (i.e., on the order of the Earth radius), and our universe appears from outside as a particle (bottleneck, see Fig. 2) having a very small rest mass $m_0 \approx 2 \times 10^{-37}$ g and a radius of curvature of $R_h \approx 10^{-3}$ cm.

It is interesting to note that, if we assume that the internal radius and the total mass of the observed universe are two orders of magnitude higher than the value estimated from the Hubble constant and apply relations (20) to charges with parameters of the known objects (maximon, electron, and universe), we will readily find that they form a power series,

$$\xi = q^n,$$

where $q = 4.52 \times 10^{10}$. Then one more "particle" (with $n = 1$), which can be conditionally called the "mifion," can exist between the maximon ($n = 0$) and the electron ($n = 2$). The electron is followed by the universe ($n = 3$) (see table).

Spectrum of electric charge parameters

| Parameters | | | External | | Internal | | |
|---|---|---|---|---|---|---|---|
| n | Object | $\xi = q^n$ | $R_h$, cm | $m_0$, g | $R_{max}$, cm | $M$, g | $\rho_0$, g/cm$^3$ |
| 0 | Maximon | 1 | $1.38 \times 10^{-34}$ | $1.86 \times 10^{-6}$ | $1.38 \times 10^{-34}$ | $1.86 \times 10^{-6}$ | $1.69 \times 10^{95}$ |
| 1 | Mifion | $4.52 \times 10^{10}$ | $0.62 \times 10^{-23}$ | $4.11 \times 10^{-17}$ | $1.99 \times 10^{-13}$ | $1.34 \times 10^{15}$ | $4.06 \times 10^{52}$ |
| 2 | Electron | $2.04 \times 10^{21}$ | $2.82 \times 10^{-13}$ | $0.91 \times 10^{-27}$ | $4.06 \times 10^{8}$ | $2.73 \times 10^{36}$ | $0.98 \times 10^{10}$ |
| 3 | Universe | $0.92 \times 10^{32}$ | $1.27 \times 10^{-2}$ | $2.02 \times 10^{-38}$ | $0.83 \times 10^{30}$ | $0.56 \times 10^{58}$ | $2.35 \times 10^{-33}$ |

## 10. COSMOLOGICAL CONSEQUENCE

The fact that the internal radius and mass of the universe can be larger than those estimated in astrophysics in the approximation of the invariability of the Hubble constant $h$ and from the linear Doppler effect does not contradict the Einstein-Friedman cosmology [1]: over long distances and time intervals, we must take into account the fact that "radius" $a(\eta)$ does not remain constant during light propagation in an expanding hypersphere, but increases, and that the relationship of the observed frequency of light with the radiation frequency and the velocity of the source is determined not by the linear Lorentz transformations (in which the velocity is limited to the velocity of light), but by general covariant transformations.

The relation between the "recession velocity" $\beta = \chi \, d\ln a/d\eta$ and the "red shift" in the homogeneous model disregarding the charge (in the Friedmann–Tolman metric) has the form [1]

$$\frac{\omega}{\omega_0} = \left( \sin\frac{\eta - \chi}{2} / \sin\frac{\eta}{2} \right)^2,$$

$$\beta = hl/c = \chi \, ctg\frac{\eta}{2}, \quad (21)$$

$$h = \frac{c}{a^2}\frac{da}{d\eta} = \frac{c}{2a_0}\frac{\cos(\eta/2)}{\sin^3(\eta/2)}.$$

These relations show that the Hubble constant $h$ and the recession velocity $\beta$ turn to infinity at singularity ($\eta = 0$) and are zero at the maximal expansion state ($\eta = \pi$). This does not lead to any contradiction since $\beta$ in the present case is not a physical velocity: all points of the hypersphere are at rest relative to it, and the space itself expands.

## 11. EXPERIMENTAL CONFIRMATIONS

The assumption that the charge of the Universe is equal to $e$ does not contradict the astrophysical data indicating the presence of a nonvanishing mean electromagnetic field in it. It is this field, which is strong enough in a state close to maximal compression, that ensures acceleration of relict particles generated ("tempered") at this stage of evolution of the universe to ultrahigh energies ($10^{21}$–$10^{22}$ eV) [2]. In the present state, which is close to maximal expansion, these particles become primary sources of experimentally observed extensive air showers (EAS).

It is difficult to explain the presence of such particles in cosmic noise by other mechanisms (stellar or galactic). A particle may accumulate an ultrarelativistic energy of $10^{22}$ eV on the scale of the Universe. Due to radiation losses during accelerated motion along a curvilinear trajectory, this energy is lower than the critical energy $ec^2/\sqrt{k}$ (the maximon rest energy), which is on the order of $10^{28}$ eV [2].

Further, a rigorous result of the theory is that a charge with the parameters of the electron contains (in the case of a uniform density of the inner world in the initial state) a universe with a mass on the order of $10^3$ masses of the Sun. This can also be verified experimentally if we assume that bursts of Supernovas observed in astrophysics can be interpreted not as a result of a catastrophic collapse of burnt-out cooled stars, but as the release of a part of the internal energy by an elementary particle that loses its stability for one reason or other (say, as a result of insufficiently considered experiments on accelerators or due to civilizations existing on planets of the given galaxies).

## 12. CONCLUSIONS

First integrals of the Einstein-Maxwell equations for the system under study were obtained by Markov and Frolov in 1972 [3]. In [2], these equations were integrated completely. The exact solution of the Einstein-Maxwell equations for the inner space of a spherically symmetric electric charge [2, 4] implies that

(i) the electric charge is a gravitational object in the GTR; the radius of the Gaussian curvature of the bottleneck connecting the inner nonstationary pulsating semi-confined world of dust and the outer Reissner–Nordstrem vacuum world is equal to doubled classical radius; (ii) the space curvature (gravitational field) removes the Coulomb divergence of the field produced by a point charge in a 2D Minkowski space-time; the radial electric field in the co-moving reference frame attains its maximal value on two parallel bottlenecks and decreases in inverse proportion to the squared radius of the Gaussian curvature with increasing distance from the bottlenecks to the bulk of the charge and to vacuum; (iii) physical constants $e$ and $m_0$ are first integrals of the Einstein–Maxwell equations; all physical parameters (electric charge, rest mass, radial electric field, and dust density) can be expressed in terms of the curvatures of the 4-space and can be determined by measuring the curvatures at any point of the space; (iv) the rest mass is the total (gravitational and observed) mass of the inner world on the bottleneck; the smallness of the gravitational radius as compared to the classical radius of many "elementary" particles indicates not a negligibly small role of gravitational effects over the classical length, but rather strong "gravitational mass defect" of the inner world on the bottleneck due to focusing (attracting) action of the gravitational field, when the large space curvature reduces the observed mass of the object; for a charge with the electron parameters, the value of $\xi \gg 1$:

$e \gg \sqrt{k} m_0$; i.e., the electric charge is much larger than the "gravitational charge"; (v) an elementary particle (electron, proton, maximon, etc. in the "nonquantum" representation of the GTR) and the Universe are formally a single object considered from outside and inside; i.e., the micro- and macroworlds are identical; for uniform initial conditions, the "electron" contains a universe with a mass on the order of $10^3$ masses of the Sun and a maximal radius on the order of the Earth radius, while our Universe (if its charge is $e$) appears from outside as a particle whose mass is on the order of $10^{-37}$–$10^{-38}$ g and a bottleneck radius on the order of $10^{-3}$–$10^{-2}$ cm; (vi) in the given problem, the global space-time is topologically nontrivial and "layered" (this should not be confused with stratification in the gauge field theory); various objects (electron, universe, etc.) are tunnels connecting these parallel layers; this solution theoretically confirms the correctness of the ideas concerning the neighborhood of the GTP (e.g., Wheeler's mole burrows [5]); if an "electron" emerges in the world (the bottleneck corresponding to a negative charge), a "positron" emerges in the parallel space (the bottleneck with a positive charge), which is an antiworld; thus, a world of particles and an antiworld of charged antiparticles are located on two parallel orientable 3-hypersurfaces; (vii) since a charge particle consists of dust, which in turn is formed by charged particles (burrows between the layers of vacuum spaces), the space as a whole can be supplied with a nontrivial topological structure of a closed set that is not dense anywhere (an everywhere "perforated" Cantor-type set); consequently, the general problem of existence might have a paradoxical solution: there exists something equal to zero (having zero measure).

The problem considered here shows that the Einstein GTR can be geometrized: the electromagnetic field and matter have a gravitational (geometric) mapping. This statement is of methodological importance: the prevailing opinion that gravitation is a field equivalent in properties to other physical fields and having only a geometrical interpretation like other fields is archaic. On the contrary, the gravitational field has a synthesizing meaning: any physical field possessing an energy-momentum tensor can be mapped on the geometry of space whose curvature is precisely the gravitational field. Matter is equivalent to a gravitational field, which is equivalent to a curved space. This is the essence of the Einstein equations proper and of the rigorous GTR principle of equivalence. All of other formulations of this principle (equality of the inertial and gravitational masses, local "vanishability" of gravitational field, local "rectifiability" of space, etc.) are of limited (nonrelativistic) nature.

The prevailing idea that the gravitational field is significant either over limiting lengths of the Planck type (which is approximately an order of magnitude larger than the critical radius) in microworld or on the scale of the Universe in the megaworld is refined in the GTR: gravitational fields are manifested over any length as the maps of physical fields onto space-time geometry. Thus, a gravitational field cannot be "weaker" than, say, an electromagnetic field since it is precisely this field in the form of tensions of the curved space.

It should also be noted that the traditional concept on a gravitational field as a "classical" field can also be revised in near future. The GTR sets no intrinsic limitations on the values of any parameters. Conversely, a nongravitational field should be subjected to quantization, while the origin of discrete quantum effects should be explained with the help of a "continuous" gravitational field (it was mentioned above that this formed the initial Einstein program).

ACKNOWLEDGMENTS

The author is obliged to mention Ya.B. Zeldovich, who once showed his interest in the problem, and to thank N.V. Mitskevich, I.D. Novikov, and V.P. Frolov for their remarks made in the 1970-80s. Thanks are also due to Yu.S. Vladimirov, V.N. Mel'nikov, and participants of their seminars (especially K.A. Bronnikov); participants of the seminar carried out at the Moscow Physical Engineering Institute (especially N.S. Trushkin), and also E.D. Zhizhin and A.V. Berkov for discussions and valuable remarks. The author is grateful to B.Yu. Bogdanovich, B.N. Onykiĭ, and É.Ya. Shkolnikov for their support and to M.Yu Lukashin, L.A. Sukhanova and A.Yu. Khlestkov for his help in this research.

APPENDIX

A Brief Review of Publications on Nontrivial Geometrical Structures in the Theory of Gravitation

Obtaining regular solutions to the Einstein equations with a nontrivial topology (black holes, bottlenecks, burrows, tubes, bubbles, etc.) have been made for a long time and certain advances have been made in this direction (see, for example, [6-15]). However, the general conclusion that has been drawn is distressing [6, 9]: a space with a bottleneck (horn, wormhole) and two asymptotically-planar worlds can be constructed only in "pathological" cases from the standpoint of generally accepted concept of causality (negative energy density of the matter generating a gravitational field, violation of the weak energy condition [16], etc.).

Nevertheless, we have obtained a solution that describes the internal structure of an electric charge without any singularities (geometrical or physical) on the bottleneck. This can be explained by the difference in the formulation of the problem. In previous works, static ($\dot{R} = 0$) vacuum ($\varepsilon_s = 0$, $\rho_f = 0$) worlds with a scalar field, as well as worlds with an electric and scalar field with different Lagrangians were considered in the framework of spherical symmetry [6-15]. In most cases, metric (1) of these worlds can be reduced to the form [6] $\nu(r) + \lambda(r) = 0$. For such worlds, the right-hand side of the Einstein equations implies that the difference between two mixed components of the conservative Einstein tensor must be

$$G_0^0 = G_1^1 = \kappa(\varepsilon_s + p_s);$$

on the other hand, it follows from the left-hand side of the Einstein equations that this difference is given by

$$-e^{-\lambda(r)} R'' / R.$$

In [6-15], the bottleneck was defined as an infinitely long 3D tube with a finite radius of curvature $R(r)$, which is minimal on the 2-surface $r = r_h$, i.e.,

$$R'_h = 0, \quad R''_h > 0.$$

It was also assumed that metric coefficient $e^{\nu_h}$ on this surface has a finite value. It turns out that these conditions can be satisfied only for a negative sum of the energy density and pressure of any matter.

Worlds whose source was supplemented with the so-called cosmological Λ term were considered. If the latter term is identified as the first type of the energy-momentum tensor according to Petrov [1], we can consider a matter with an energy density Λ/κ and isotropic pressure −Λ/κ as an exotic matter (i.e., the matter with an ultrarelativistic equation of state, for which the sign of energy density is opposite to the sign of pressure. This is usually regarded as pathology both for Λ > 0 and for Λ < 0.

Here, we managed to avoid this owing to another formulation of the problem: we considered a class of nonstationary metrics (1),

$\dot R \neq 0$, generated by a dustlike neutral matter and an electromagnetic field, which is represented by a radial electric field in the reference frame co-moving with the dust.

Now, the left-hand side of the Einstein equations appears quite differently. It does not contain the second derivative of radius $R$ of the Gaussian curvature with respect to coordinate $r$ and, by virtue of the solution to the Einstein–Maxwell equations, is identically equal to the positive right-hand side:

$$G_0^0 - G_1^1 = 2\left(1 - f(r)^2 + \dot R^2\right)'/R^{2'} - 2\ddot R/R = \kappa \varepsilon_s > 0.$$

This expression (as well as other physical and geometrical quantities) remains finite on the bottleneck. If we define the bottleneck as a 2-surface on which $R'_h = 0$, $\dot R_h = 0$ (this does not exhaust all possible definitions of the bottleneck), then $e^{\lambda_h} = 0$ for $\xi > 1$; i.e., only a purely coordinate singularity exists on it (the spherical system of coordinates degenerates).